\documentclass[twocolumn,showpacs,preprintnumbers,amsmath,amssymb]{revtex4}

\usepackage{graphicx}
\usepackage{dcolumn}
\usepackage{bm}

\begin{document}

\title{Continuous extremal optimization for Lennard-Jones Clusters}
\author{Tao Zhou$^{1}$}
\email{zhutou@ustc.edu}
\author{Wen-Jie Bai$^{2}$}
\author{Long-jiu Cheng$^{2}$}
\author{Bing-Hong Wang$^{1}$}
\email{bhwang@ustc.edu.cn}
\affiliation{%
$^{1}$Nonlinear Science Center and Department of Modern Physics,\\
University of Science and Technology of China, \\
Hefei Anhui, 230026, PR China\\
$^{2}$Department of Chemistry, University of Science and
Technology of China, \\
Hefei Anhui, 230026, PR China\\
}%

\date{\today}

\begin{abstract}
In this paper, we explore a general-purpose heuristic algorithm
for finding high-quality solutions to continuous optimization
problems. The method, called continuous extremal
optimization(CEO), can be considered as an extension of extremal
optimization(EO) and is consisted of two components, one is with
responsibility for global searching and the other is with
responsibility for local searching. With only one adjustable
parameter, the CEO's performance proves competitive with more
elaborate stochastic optimization procedures. We demonstrate it on
a well known continuous optimization problem: the Lennerd-Jones
clusters optimization problem.

\end{abstract}


\maketitle

\section{Introduction}
The optimization of system with many degrees of freedom with
respect to some cost function is a frequently encountered task in
physics and beyond. One special class of algorithms used for
finding the high-quality solutions to those NP-hard optimization
problems is the so-called nature inspired algorithms, including
simulated annealing(SA)\cite{SA1,SA2}, genetic
algorithms(GA)\cite{GA1,GA2,GA3}, genetic
programming(GP)\cite{GP}, and so on.

In recent years, a novel nature inspired algorithm named extremal
optimization(EO) is proposed by Boettcher and
Percus\cite{EO1,EO2,EO3,EO4,EO5}, which is very sententious and
competitive comparing with some well known algorithms like SA, GA,
GP etc.. To make the underlying mechanism of EO more concrete,
let's focus on the natural selection of biological system. In
nature, highly specialized, complex structure often emerge when
their most inefficient elements are selectively driven to
extinction. For example, evolution progresses by selecting against
the few most poorly adapted species, rather than by expressly
breeding those species best adapted to their environment. The
principle that the least fit elements are progressively eliminated
has been applied successfully in the Bak-Sneppen
model\cite{BS1,BS2}, where each individual corresponding a certain
species is characterized by a fitness value, and the least fit one
with smallest fitness value and its closest dependent species are
successively selected for adaptive changes. The extremal
optimization algorithm draws upon the Bak-Sneppen mechanism,
yielding a dynamic optimization procedure free of selection
parameters.

Here we consider a general optimization problem, where the system
consists of $N$ elements, and we wish to minimize the cost
function $C(S)$ depending on the system configuration $S$. The EO
algorithm proceeds as follows:\\
(1) Choose an initial configuration $S$ of the system at will; set
$S_{\texttt{best}}:=S$.\\
(2) Evaluate the fitness value $f_i$ for each individual $i$ and
rank each individual according to its fitness value so as to the
least fit one is in the top. Use $k_i$ to denote the individual
$i$'s rank, clearly, the least fit one is of rank 1. Choose one
individual $j$ that will be changed with the probability $P(k_j)$,
and then, only randomly change the state of $j$ and keep all other
individuals' state unaltered. Accept the new configuration $S'$
unconditionally $S:=S'$, and if $C(S)<C(S_{\texttt{best}})$ then
set $S_{\texttt{best}}:=S$. \\
(3) Repeat at step (2) as long as desired. \\
(4) Return $S_{\texttt{best}}$ and $C(S_{\texttt{best}})$.

The efficiency of EO algorithm is sensitive to the probability
function $P(k)$. In basic EO, $P(1)=1$ and for any $k(2\leq k\leq
N)$, $P(k)=0$. A more efficient algorithm, the so-called
$\tau$-EO, can be obtained through a slight modification from
basic EO. In $\tau$-EO, $P(k)\sim k^{-\tau}$ where $\tau>0$. Of
course, aiming at idiographic optimization problems, one can
design various forms of $P(k)$ to improve the performance of basic
EO. For example, Middleton has proposed the jaded extremal
optimization(JEO) method for Ising spin glass system by reducing
the probability of flipping previously selected spins, which
remarkably improved the efficiency of EO\cite{EO6}.

The previous studies indicate that EO algorithm can often
outperform some far more complicated or finely tuned algorithm,
such as SA or GA, on some famous NP-hard\cite{NP} discrete
optimization problems, including graph
partitioning\cite{EO1,EO2,EO7}, travelling salesman
problem\cite{EO1}, three-coloring problem\cite{EO8,EO9}, finding
lowest energy configuration for Iring spin glass
system\cite{EO6,EO8,EO10}, and so on. However, many practical
problems can not be abstracted to discrete form, thus to
investigate EO's efficiency on continuous optimization
problems\cite{EX1} is not only of theoretic interest, but also of
prominent practical worthiness.

In this paper, a so-called continuous extremal optimization(CEO)
algorithm aiming at continuous optimization problem will be
introduced, which can be considered as a mixing algorithm
consisting of two components, one is with responsibility for
global searching and the other is with responsibility for local
searching. With only one adjustable parameter, the CEO's
performance proves competitive with more elaborate stochastic
optimization procedures. We demonstrate it on a well known
continuous optimization problem: the Lennerd-Jones(LJ) clusters
optimization problem.

This paper is organized as follows: in section 2, the LJ clusters
optimization problem will be briefly introduced. In section 3, we
will give the algorithm proceeds of CEO. Next, we give the
computing results about the performance of CEO on LJ clusters
optimization problem. Finally, in section 5, the conclusion is
drawn and the relevances of the CEO to the real-life problems are
discussed.

\section{Lennerd-Jones Clusters Optimization Problem }

Continuous optimization problem is ubiquitous in materials
science: many situation involve finding the structure of clusters
and the dependence of structure on size is particularly complex
and intriguing. In practice, we usually choose a potential
function to take the most steady structure since it's considered
to be in possession of the minima energy. However, in all but the
simplest cases, these problem are complicated due to the presence
of many local minima. Such problem is encountered in many area of
science and engineering, for example, the notorious protein
folding problem\cite{lj1}.

\begin{figure}
\scalebox{0.75}[0.8]{\includegraphics{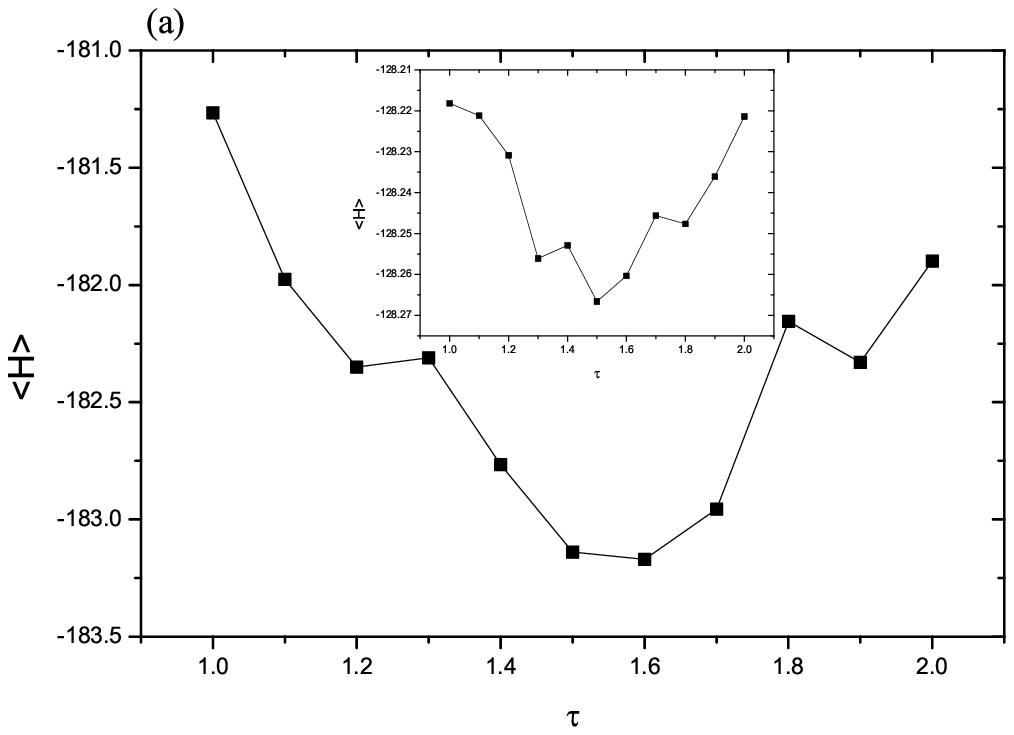}}
\scalebox{0.8}[0.9]{\includegraphics{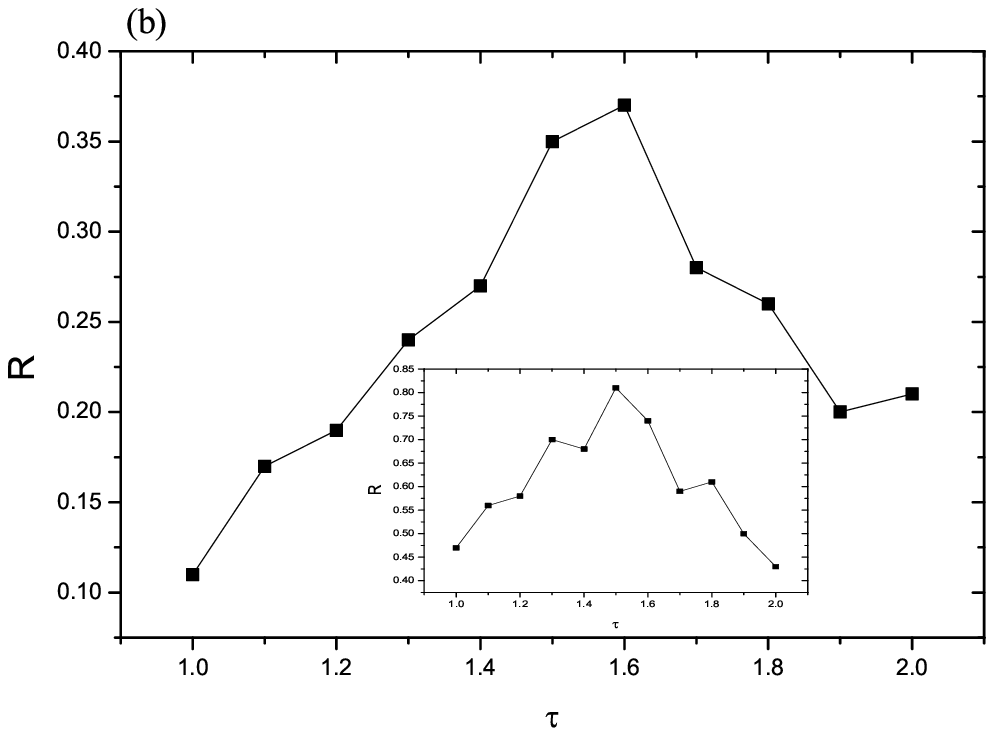}}
\caption{\label{fig:epsart} The details of $\tau$-CEO for $\tau
\in [1,2]$. Figure 1{\bf a} shows the average energies obtained by
CEO over 200 runs, and figure 1{\bf b} exhibits the success rate
of hitting the global minima in 200 runs\cite{EX2}. For both
figure 1{\bf a} and 1{\bf b}, the main plot and inset represent
the case $N=40$ and $N=30$ respectively. One can find that, the
best $\tau$ corresponding lowest average energy and highest
success rate is approximate to 1.5.}
\end{figure}

As one of the simplest models that exhibits such behavior
\cite{lj2} one may consider the problem of finding the
ground-state structure of nanocluster of atoms interacting through
a classical Lennerd-Jones pair potential, in reduced units given
by
\begin{equation}
V(r)={\frac{1}{r^{12}}-\frac{1}{r^{6}}}
\end{equation}
where $r$ is the distance between two atoms. This potential has a
single minimum at $r=\sqrt[6]{2}$, which is the equilibrium
distance of two atoms. It can, of course, easily be reduced to an
arbitrary LJ-potential by a simple rescaling of length and energy
units. The $i$th atom has energy
\begin{equation}
\emph{E}_{i}=\frac{1}{2}\sum_{j\neq i}\emph{V}(\emph{r}_{ij})
\end{equation}
and the total energy for \emph{N} atoms is
\begin{equation}
\emph{E}=\sum_{i}\emph{E}_{i}
\end{equation}
The optimization task is to find the configuration with minimum
total potential energy of a system of $N$ atoms, each pair
interacting by potential of the form (1). Clearly, a trivial lower
bound for the total energy is $-N(N-1)/2$, obtained when one
assumes that all pairs are at their equilibrium separation. For
$N=2,3,4$ the lower bound can actually be obtained in
three-dimensional space, corresponding respectively to a dimer,
equilateral triangle, and regular tetrahedron, with all
interatomic distance equal to 1. However, from $N=5$ onwards it is
not possible to place all the atoms simultaneously at the
potential minimum of all others and the ground-state energy is
strictly larger than the trivial lower bound. This system has been
studied intensely \cite{lj3} and is known to have an exponential
increasing number of local minima, growing roughly as
$e^{0.36N+0.03N^2}$ near $N=13$, at which point there are already
at least 988 minima \cite{lj3}. If this scaling continues, more
than $10^{140}$ local minima exist when $N$ approach 100.

\section{Continuous extremal optimization}
The continuous extremal optimization algorithm is consisted of two
components, one is the classical EO algorithm with responsibility
for global searching, and the other is a certain local searching
algorithm. We give the general form of CEO algorithm by way of the
LJ clusters optimization problem as follows:\\
(1) Choose an initial state of the system, where all the atoms are
placed within a spherical container with
radius\cite{radius1,radius2}
\begin{equation}
radius=r_e[\frac{1}{2}+(\frac{3N}{4\pi \sqrt{2}})^{1/3}],
\end{equation}
where $r_e=\sqrt[6]{2}$ is the equilibrium distance and $N$
denotes the number of atoms. Set the minimal energy $E_{\texttt{min}}=0$.\\
(2) Use a certain local searching algorithm to find the local
minimum from the current configuration of system. If the local minimal energy is lower than $E_{\texttt{min}}$, then replace $E_{\texttt{min}}$ by the present local minimum. \\
(3) Rank each atom according to its energy obtained by Equ.(2).
Here, the atom who has highest energy is the least fit one and is
arranged in the top of the queue. Choose one atom $j$ that will be
changed with the probability $P(k_j)$ where $k_j$ denotes the rank
of atom $j$, and then, only randomly change the coordinates of $j$
and keep all other atoms' positions unaltered. Accept the new
configuration unconditionally. \\
(4) Repeat at step (2) as long as desired. \\
(5) Return the minimal energy $E_{\texttt{min}}$ and the
corresponding configuration.

For an idiographic problem, one can attempt various local
searching algorithms and pitch on the best one. In this paper, for
the LJ clusters optimization problem, we choose limited memory
BFGS method(LBFGS) qua the local searching algorithm. The BFGS
method is an optimization technique based on quasi-Newton method
proposed by Broyden, Fletcher, Goldfard and Shanno. LBFGS proposed
by Liu and Nocedal\cite{radius2,LBFGS} is especially effective on
problems involving a large number of variables. In this method, an
approximation $H_k$ to the inverse of the Hessian is obtained by
applying $M$ BFGS updates to a diagonal matrix $H_0$, using
information from the previous $M$ steps. The number $M$ determines
the amount of storage required by the routine, which is specified
by the user, usually $3\leq M\leq 7$ and in our computation $M$ is
fixed as 4.

\section{Computing results}

\begin{figure}
\scalebox{0.75}[0.8]{\includegraphics{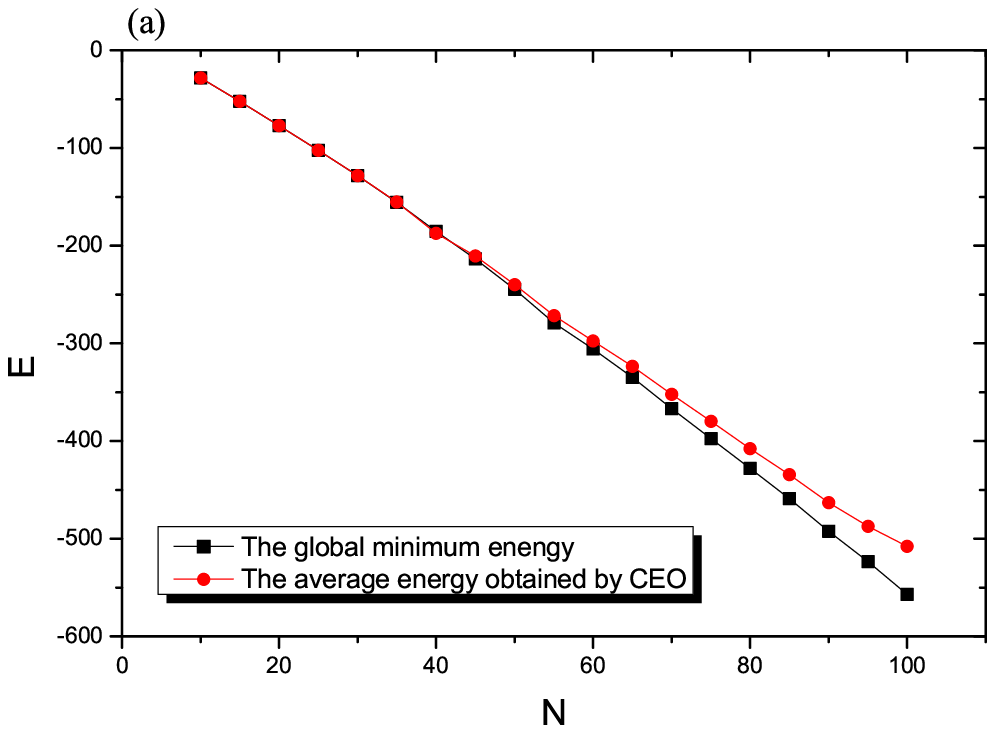}}
\scalebox{0.8}[0.9]{\includegraphics{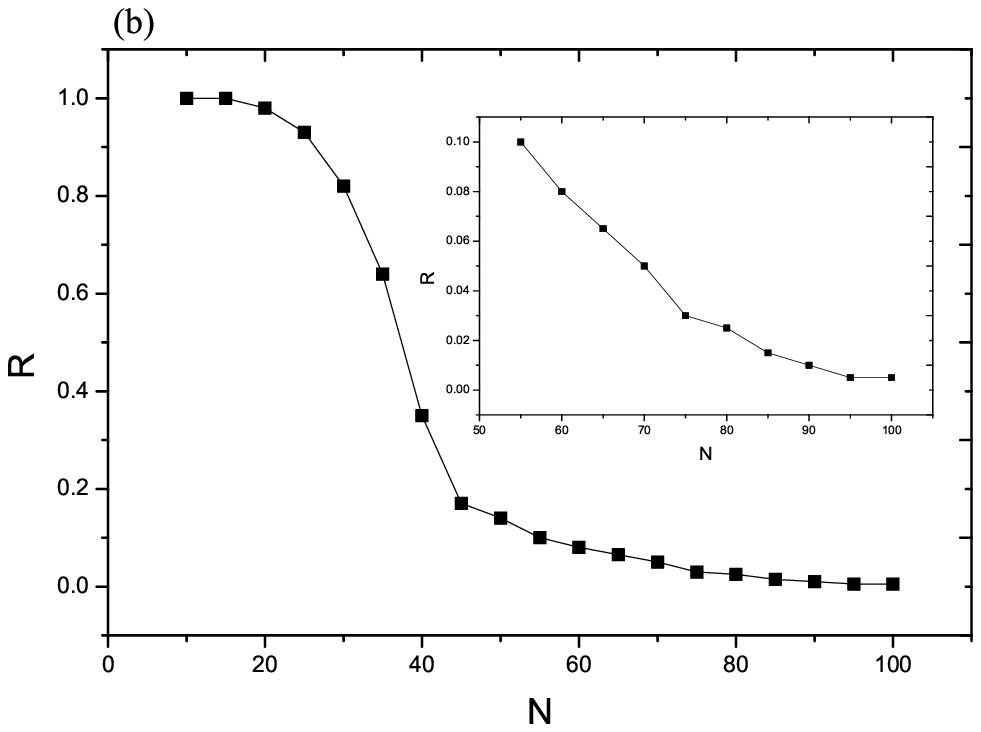}}
\caption{\label{fig:epsart} The performance of CEO algorithm on LJ
clusters optimization problem. In figure 2{\bf a}, the red circles
represent the average energies obtained by CEO over 200 runs,
where the black squares represent the global minima. Figure 2{\bf
b} shows the success rate of hitting the global minima in 200
runs, the inset is the success rate for $N>50$ that may be unclear
in the main plot.}
\end{figure}

Similar to $\tau$-EO, we use $\tau$-CEO algorithm for the LJ
clusters optimization problem, where the probability function of
CEO is $P(k)\sim k^{-\tau}$. Since there are $\frac{N^2}{2}$ pairs
of interactional atoms in a LJ cluster of size $N$, we require
$\alpha N^2$ updates where $\alpha$ is a constant and fixed as 100
in the following computation. In order to avoid falling into the
same local minimum too many times, before running LBFGS, we should
make the system configuration far away from last local minimum,
thus we run LBFGS every 20 time steps. That is to say, for a LJ
cluster of size $N$, the present algorithm runs EO $100N^2$ times
and LBFGS $5N^2$ times in total.

We have carried out the $\tau$-CEO algorithm so many times for
different $\tau$ and $N$, and find that the algorithm performs
better when $\tau$ is in the interval [1,2]. In figure 1, we
report the details for $1\leq \tau \leq 2$, where figure 1{\bf a}
shows the average energies obtained by CEO over 200 runs, and
figure 1{\bf b} exhibits the success rate $R$ of hitting the
global minima\cite{EX2}. For both figure 1{\bf a} and 1{\bf b},
the main plot and inset represent the case $N=40$ and $N=30$
respectively. The readers should note that, although the
difference of average energies between two different $\tau$ is
great in the plot, it is very very small in fact. For the case
$N=40$, the best value of $\tau$ is $\tau=1.6$ corresponding the
lowest average energy and highest success rate, however, the
performance of CEO for $\tau=1.5$ is almost the same as $\tau=1.6$
in this case but obviously better than $\tau=1.6$ in the case
$N=30$. Therefore, in the following computation, we set
$\tau=1.5$. We have also compared the performance of CEO on larger
LJ clusters for $\tau=1.5$ and $\tau=1.6$, the two cases are
pretty much the same thing and $\tau=1.5$ is a little better.

\begin{figure}
\scalebox{0.75}[0.8]{\includegraphics{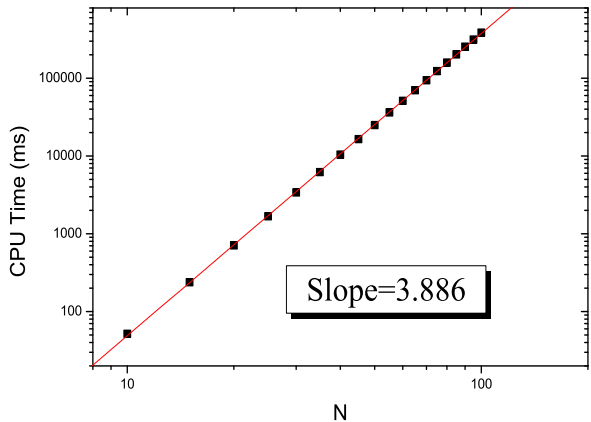}}
\caption{\label{fig:epsart} The average CPU time(ms) over 200 runs
va the size of LJ clusters. In the log-log plot, the data can be
well fitted by a straight line with slope $3.886\pm 0.008$, which
indicates that the increasing tendency of CPU time $T$ vs cluster
size is approximate to power-law form as $T\sim N^{3.886}$.}
\end{figure}

We demonstrate that for all the LJ clusters of size $N$ not more
than 100, the global minima can be obtained by using CEO
algorithm. In figure 2, we report the performance of CEO on LJ
clusters optimization problem according to 200 independent runs.
In figure 2{\bf a}, the red circles represent the average energies
obtained by CEO over 200 runs, where the black squares represent
the global minima. One can see that the deviation from global
minimum becomes greater and greater when the cluster size getting
larger and larger, which indicates that for very large LJ cluster,
CEO may be a poor algorithm. Figure 2{\bf b} shows the success
rate of hitting the global minima in 200 runs, the inset is the
success rate for $N>50$ that may be unclear in the main plot. For
both the case $N=95$ and $N=100$, the global optimal solutions
appears only once in 200 runs.

Although CEO is not a all-powerful algorithm and it may perform
poorly for very large LJ clusters, we demonstrate that it is
competitive or even superior over some more elaborate stochastic
optimization procedures like SA\cite{SAcom}, GA\cite{GAcom} in
finding the most stable structure of LJ clusters with minimal
energy.

Finally, we investigate the average CPU time over 200 runs vs the
size of LJ clusters. The computations were carried out in a single
PentiumIII processor(1GHZ). From figure 3, in the log-log plot,
the data can be well fitted by a straight line with slope
$3.326\pm 0.008$, which indicates that the increasing tendency of
CPU time $T$ vs cluster size is approximate to power-law form as
$T\sim N^{3.886}$. That means the CEO is a polynomial algorithm of
order $O(N^4)$.

\section{Conclusion and Discussion}

In this paper, we explore a general-purpose heuristic algorithm
for finding high-quality solutions to continuous optimization
problems. The computing results indicate that this simple approach
is competitive and sometimes can outperform some far more
complicated or finely tuned nature inspired algorithm including
simulated annealing and genetic algorithm, on a well-known NP-hard
continuous optimization problem for LJ clusters(see
reference\cite{SAcom,GAcom} for comparison). According to EO's
updating rule, it is clear that EO has very high ability in global
searching, thus to combine EO and a strong local searching
algorithm may produce a high efficient algorithm for continuous
optimization problems.

Recently, several novel algorithms aiming at LJ clusters
optimization problem have been proposed, such as fast annealing
evolutionary algorithm\cite{radius2}, conformational space
annealing method\cite{CSA}, adaptive immune optimization
algorithm\cite{AIOA}, cluster similarity checking
method\cite{CSC}, and so forth. These algorithms consider more
about the special information about LJ clusters and perform better
than CEO. However, we have not found a compellent evidence
indicating that there exists a general-purpose algorithm like SA
or GA entirely preponderate over CEO on LJ cluster optimization
problem. It is worthwhile to emphasize that, in this paper, we do
not want to prove that the CEO is an all-powerful algorithm, even
do not want to say that the CEO is a good choice for chemists on
LJ cluster optimization problem since a general-purpose method
often perform poorer than some special methods aiming at an
idiographic problem. The only thing we want to say is the CEO, an
extension of nature inspired algorithm EO, is a competitive
algorithm and needs more attention.

Further more, to demonstrate the efficiency of CEO, much more
experiments on various hard continuous optimization problems
should be achieved.

\begin{acknowledgments}

This work is supported by the outstanding youth fund from the
National Natural Science Foundation of China (NNSFC) under Grant
No. 20325517, and the Teaching and Research Award Program for
Outstanding Young Teacher (TRAPOYT) in higher education
institutions of the Ministry of Education (MOE) of China, the
State Key Development Programme of Basic Research of China (973
Project), the NNSFC under Grant No. 10472116, 70471033 and
70271070, and the Specialized Research Fund for the Doctoral
Program of Higher Education (SRFDP No.20020358009).

\end{acknowledgments}


\begin{thebibliography} {SA1}

\bibitem{SA1} S. Kirkpatrick, C. D. Gelatt and M. P. Vecchi,
Science {\bf 220}, 671(1983).
\bibitem{SA2} E. H. L. Aarts and J. H. M. Korst, {\it Simulated
annealing and Boltzmann machines} (John Wiley and Sons, 1989).
\bibitem{GA1} J. Holland, {\it Adaption in natural and artificial
systems} (The university of Michigan Press, Ann Arbor, MI, 1975).
\bibitem{GA2} D. G. Bounds, Nature {\bf 329}, 215(1987).
\bibitem{GA3} D. E. Goldberg, {\it Genetic algorithms in search,
optimization, and machine learning} (Addison-Wesly, MA, 1989).
\bibitem{GP} W. Banzhaf, P. Nordin, R. Keller and F. Francone,
{\it Genetic programming - an introduction} (Morgan Kaufmann, San
Francisco, CA, 1998).
\bibitem{EO1} S. Boettcher and A. G. Percus, Artificial Intelligence
{\bf 199}, 275(2000).
\bibitem{EO2} S. Boettcher, J. Phys. A {\bf 32}, 5201(1999).
\bibitem{EO3} S. Boettcher, Comput. Sci. Eng. {\bf 2}, 6(2000).
\bibitem{EO4} S. Boettcher, Comput. Sci. Eng. {\bf 2}, 75(2000).
\bibitem{EO5} S. Boettcher, A. G. Percus and M. Grigni, Lect.
Notes. Comput. Sci. {\bf 1917}, 447(2000).
\bibitem{BS1} P. Bak and K. Sneppen, Phys. Rev. Lett. {\bf 71},
4083(1993).
\bibitem{BS2} K. Sneppen, P. Bak, H. Flyvbjerg and M. H. Jensen,
Proc. Natl. Acad. Sci. 92, 5209(1995).
\bibitem{EO6} A. A. Middleton, Phys. Rev. E {\bf 69},
055701(R)(2004).
\bibitem{NP} M. R. Garey and D. S. Johnson, {\it Computers and
intractability: a guide to the theory of NP-completeness}
(Freeman, New York, 1979).
\bibitem{EO7} S. Boettcher and A. G. Percus, Phys. Rev. E {\bf
64}, 026114(2001).
\bibitem{EO8} S. Boettcher and A. G. Percus, Phys. Rev. Lett. {\bf
86}, 5211(2001).
\bibitem{EO9} S. Boettcher and A. G. Percus, Phys. Rev. E {\bf
69}, 066703(2004).
\bibitem{EO10} S. Boettcher and M. Grigni, J. Phys. A {\bf 35},
1109(2002).
\bibitem{EX1} In discrete optimization problem, the set of all the states of system is
denumerable, while in continuous optimization problem, the
corresponding set is a continuum.
\bibitem{lj1} T. P. Martin, T.Bergmann, H.G\"{o}hilich and T.
Lange, Chem. Phys. Lett. {\bf 172},209 (1990).
\bibitem{lj2} L. T. Wille, in: D. Stauffer(Ed.), Annual
Reviewsof Computational Physics VII, World Scientific, Singnapore,
2000.
\bibitem{lj3} M. R. Hoare, Advan. Chem. Phys. {\bf 40} (1979)
49, and references therein.
\bibitem{radius1} B. Xia, W. Cai, X. Shao, Q. Guo, B. Maigret and
Z. Pan, J. Mol. Struct.(Theochem) {\bf 546}, 33(2001).
\bibitem{radius2} W. Cai, Y. Feng, X. Shao and Z. Pan, J. Mol. Struct.(Theochem) {\bf 579}, 229(2002).
\bibitem{LBFGS} D. C. Liu and J. Nocedal, Math. Progm. B {\bf 45}, 503(1989).
\bibitem{EX2} The complete up-to-date list of the global minima of
LJ clusters can be download from the web
http://brian.ch.cam.ac.uk/.
\bibitem{SAcom} L. T. Wille, Computational Materials Science {\bf
17}, 551(2000).
\bibitem{GAcom} D. M. Deaven, N. Tit, J. R. Morris and K. M. Ho,
Chem. Phys. Lett. {\bf 256}, 195(1996).
\bibitem{CSA} J. Lee, I. -H. Lee and J. Lee, Phys. Rev. Lett. {\bf
91}, 080201(2003).
\bibitem{AIOA} X. Shao, L. Cheng and W. Cai, J. Chem. Phys. {\bf
120}, 11401(2004).
\bibitem{CSC} L. Cheng, W. Cai and X. Shao, Chem. Phys. Lett. {\bf
389}, 309(2004).

\end{thebibliography}
\end{document}